\documentclass[a4paper,12pt]{article}
\usepackage[dvipsnames]{xcolor}
\usepackage[colorinlistoftodos]{todonotes}
\usepackage{lscape}
\usepackage{longtable}
\usepackage{authblk}
\usepackage{hyperref}

\usepackage[a4paper,top=3cm,bottom=2cm,left=3cm,right=3cm,marginparwidth=1.75cm]{geometry}

\title{A decade of movement ecology}

\author[1,*]{\small Roc\'io Joo}
\author[1,2]{\small Simona Picardi}
\author[1]{\small Matthew E. Boone}
\author[3]{\small Thomas A. Clay}
\author[3]{\small Samantha C. Patrick}
\author[4]{\small Vilma S. Romero-Romero}
\author[1]{\small Mathieu Basille}

\affil[1]{\footnotesize Department of Wildlife Ecology and Conservation, Fort Lauderdale Research and Education Center, University of Florida, Fort Lauderdale, FL, USA}
\affil[2]{\footnotesize Department of Wildland Resources, Jack H. Berryman Institute, Utah State University, Logan Ut 84322, USA}
\affil[3]{\footnotesize School of Environmental Sciences, University of Liverpool, Liverpool, L69 3GP, UK}
\affil[4]{\footnotesize Universidad de Lima, Peru}
\affil[*]{Corresponding author: Rocio Joo, rocio.joo@ufl.edu}

\date{}

\begin{document}

\maketitle

\begin{abstract}
Movement is fundamental to life, shaping population dynamics,
biodiversity patterns, and ecosystem structure. Recent advances in tracking technology have enabled fundamental questions about movement to be tackled, leading to the development of the movement ecology framework (MEF), considered a milestone in the field \cite{Nathan2008}. The MEF introduced an
integrative theory of organismal movement, linking internal state, motion capacity and navigation capacity to external factors. Here, a decade later, we investigated the current state of research in the field. Using a text mining approach on \textgreater{} 8000 peer-reviewed papers in movement ecology, we explored the main research topics, evaluated the impact of the MEF, and assessed changes in the use of technological devices, software and statistical methods. The number of publications has increased considerably and there have been major technological changes in the past decade (i.e.~increased use of GPS devices, accelerometers and video cameras, and a convergence towards R), yet we found that research focuses on the same questions, specifically,
on the effect of environmental factors on movement and behavior. In practice, it appears that movement ecology research does not reflect the MEF. We call on researchers to transform the field from technology-driven to embrace interdisciplinary collaboration, in order to reveal key processes underlying movement (e.g.~navigation), as well
as evolutionary, physiological and life-history consequences of
particular strategies.
\end{abstract}

\noindent \textbf{Keywords:} movement ecology paradigm,  technology, text mining, biologging, interdisciplinarity

\newpage

\section*{Introduction: the rise of a field called movement ecology}

\begin{figure}
	\includegraphics[width=300pt]{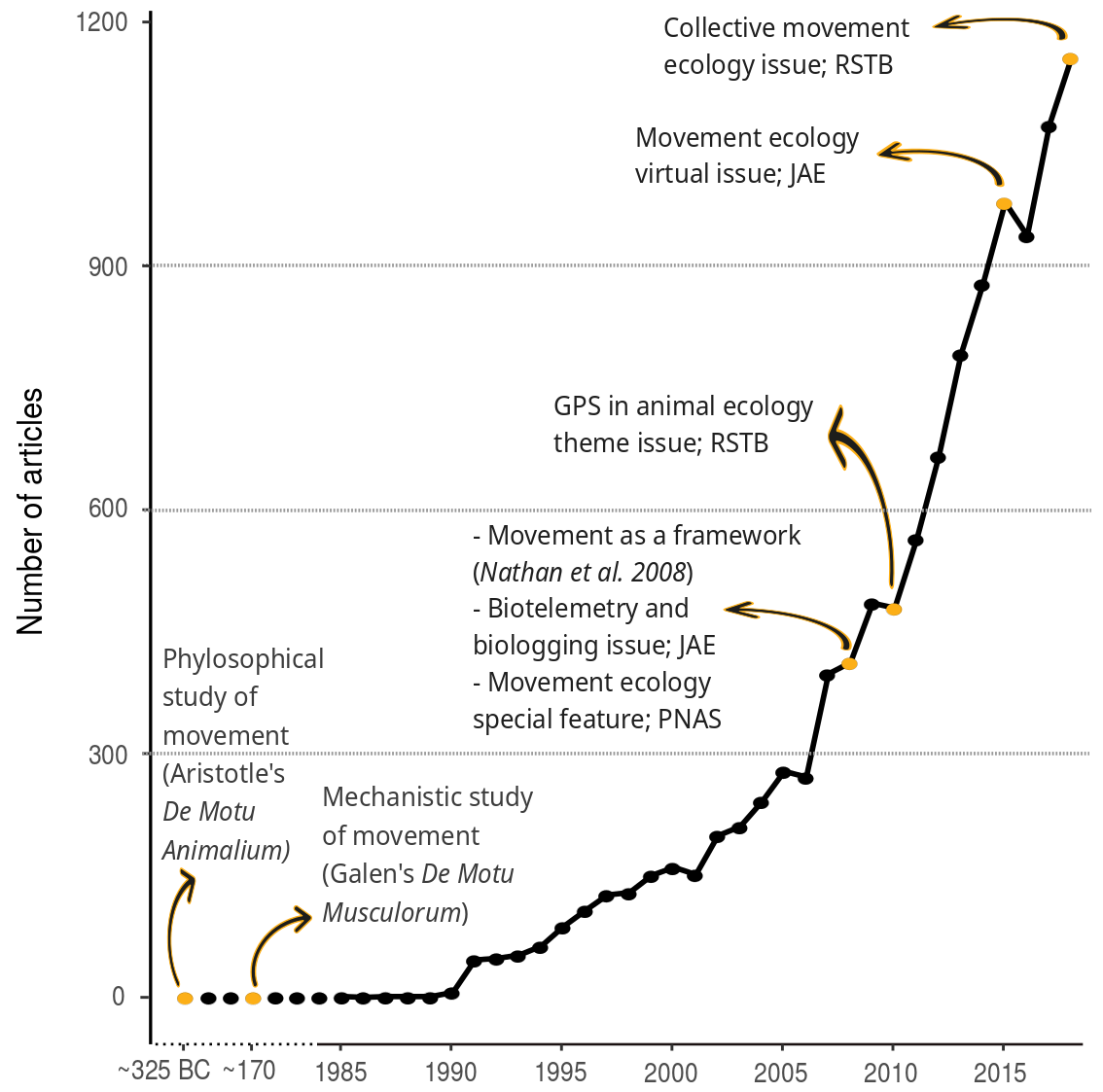}
	\caption{Timeline of movement ecology papers and milestones in the field}
	\label{fig:timeline}
\end{figure}

Movement, defined as a change in position of an individual in time, has
been studied for millennia, from philosophical (Aristotle's
\textit{De motu animalium} 384-322 BC) and mechanistic perspectives
(Galen's \textit{De motu musculorum} 129-210 AD) (Fig.
\ref{fig:timeline}), but more recently has diversified into several
research fields, such as physics \cite{Hanggi2005}, physiology \cite{Goossens2020}, data science \cite{Zheng2015},
and ecology \cite{Cooke2004}.

Around a decade ago, and as part of a PNAS' special feature on movement
ecology, a unifying conceptual framework for the study of movement was
developed \cite{Nathan2008}, aiming to promote ``the development of an integrative
theory of organism movement for better understanding the causes,
mechanisms, patterns, and consequences of all movement phenomena''. The
movement ecology framework (MEF) was born. In the same special feature,
a literature review of movement research \cite{Holyoak2008} revealed that, up to that
time, studies had mostly focused on describing movement patterns and
their links with external factors (e.g.~the environment), neglecting the
causal drivers and its consequences for individuals, populations, and
communities. It highlighted the relevance of the MEF, and encouraged
interdisciplinary work to make this possible. Technological and
methodological advances were stated as main requirements to quantify the
movement of individuals towards the study of movement in the new
integrative framework \cite{Nathan2008}.

Animal telemetry devices were first used in the mid 20th century \cite{Thums2018}.
Since then, and particularly in the last decade, loggers have become
smaller, cheaper, and more reliable, allowing for more animals to be
tagged from a large diversity of species and taxa globally, and data to
be collected at ever finer spatio-temporal resolutions \cite{Kays2015}. In the first
few decades, studies using animal telemetry commonly neglected the very
nature of the movement process: the essence of movement (i.e.~the
autocorrelation in space and time) was typically considered a nuisance
\cite{Swihart1985}, such that researchers ignored time, space, or sometimes both. More
recently and particularly in the last decade, statistical methods that
deal with space and time have become more accessible and popular in
telemetry studies (e.g.~\cite{Avgar2016},\cite{Gurarie2016},\cite{Gloaguen2015a}), allowing for statistically
sound and data-driven research on actual movement. In addition,
developments in human tracking devices have given rise to scientific
literature on human mobility. Initially inspired by animal movement
studies, human mobility science is now taking the lead in handling big
volumes of data through the development of machine learning methods for
telemetry \cite{Thums2018}.

The aforementioned changes in both ideas and technology are consistent
with an acceleration in the number of publications in movement ecology
that started about a decade ago, and coincided with a series of special
issues related to movement ecology among leading ecology journals (Fig.
\ref{fig:timeline}).

Ten years later, it is timely to ask the question: how have these
advances shaped the field of movement ecology? In this study, we
examined this research field, taking a decade-long snapshot of its
research topics, evaluating the impact of the MEF in the literature, and
assessing changes in the use of technological devices, software, and
statistical methods. We accomplished this task by reviewing
\textgreater{} 8000 movement ecology papers from the literature, using a
text mining approach. Consistent with Nathan's concept of movement
ecology, the papers considered here studied movement of organisms,
including humans. Based on a quantitative assessment, we provide an
integrative view of the state of the field, and open questions about its
future directions.

\section*{Snapshot: research topics in movement ecology}

\begin{figure*}
	\includegraphics[width=16cm]{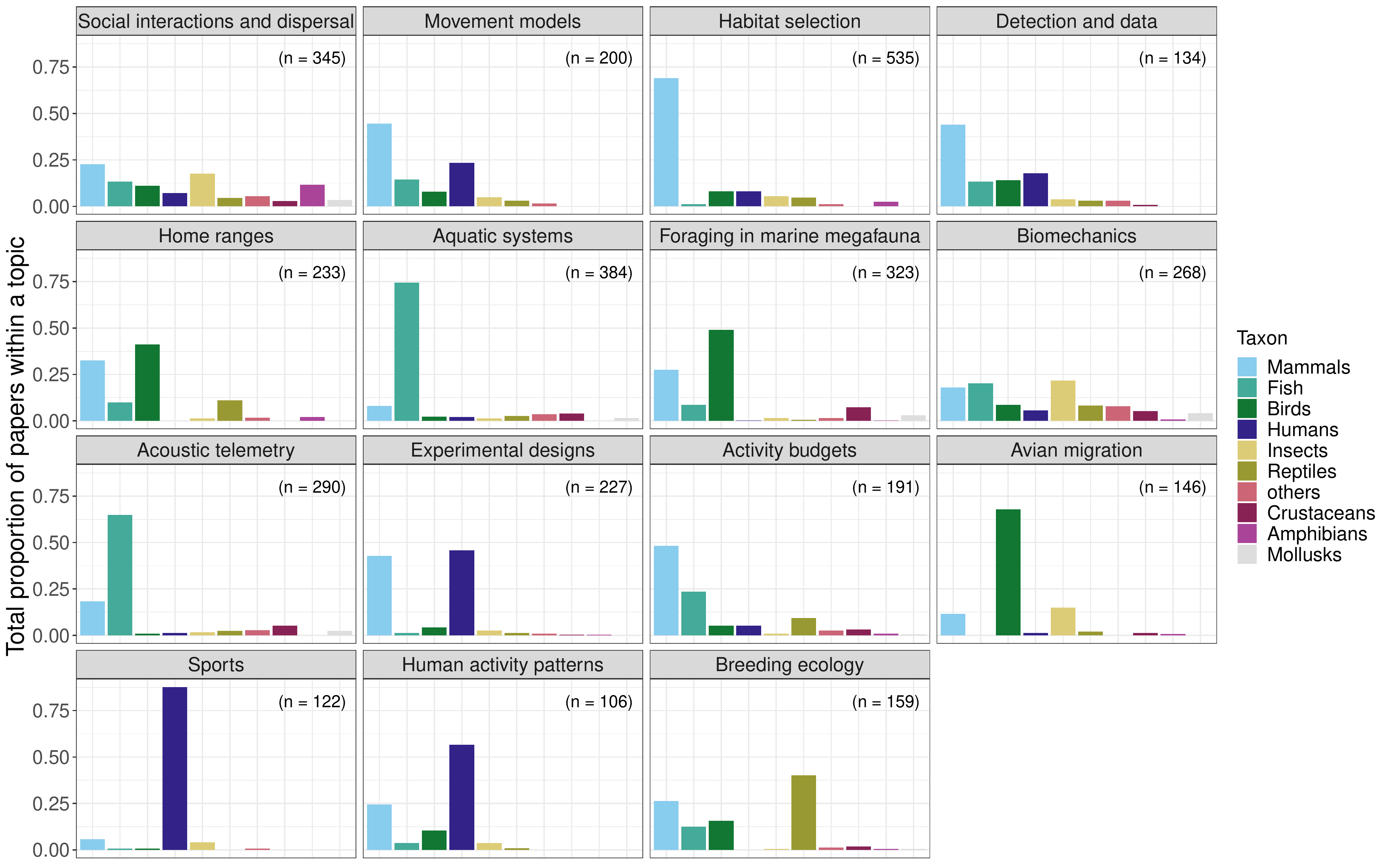}
	\caption{For each topic, relative frequencies of papers studying each
		taxonomical group. Only papers with more than $50\%$ of association to each topic ($\gamma$, see Materials and Methods) are used for this graph.}
	\label{fig:topics-taxa}
\end{figure*}

We screened all abstracts from 2009 until 2018 to identify 15 broad
topics from the words used (via Latent Dirichlet Allocation; see
Material and Methods). We chose 15 topics as a reasonable number that
would not be too large to prevent the interpretation of all topics, or
too small that the topics would be too general (see discussion in
the online Appendix section 3.1.2). These topics were, in descending order of
prevalence:

\begin{enumerate}
	\def\labelenumi{\arabic{enumi})}
	\item
	\textbf{Social interactions and dispersal}, a broad topic encompassing
	interactions with conspecifics or the environment, as well as group
	movement, changes in population density and dynamics.
	\item
	\textbf{Movement models}, encompassing any type of model
	(e.g.~generalized linear model, model selection criterion, or even
	schematical models) that could be used to study dynamics, patterns,
	and populations.
	\item
	\textbf{Habitat selection}, which encompasses choices in space use,
	influenced by resource availability or risks (e.g.~natural predators
	or human disturbance), mainly in mammalian systems (Fig.
	\ref{fig:topics-taxa}).
	\item
	\textbf{Detection and data}, focused on the collection of movement
	information and the required technological devices. This topic is also
	mostly related to mammal studies.
	\item
	\textbf{Home ranges}, mostly focused on the identification of areas
	where animals live and develop their activities, and the geographical
	extent of this area.
	\item
	\textbf{Aquatic systems}, involving the study of aquatic species,
	particularly fish, their migration, reproductive behavior and habitat,
	mostly for management purposes.
	\item
	\textbf{Foraging in marine megafauna}, consisting of foraging
	strategies and behavior of marine top predators, mostly birds and
	mammals (Fig. \ref{fig:topics-taxa}).
	\item
	\textbf{Biomechanics}, focused on body motion, swimming or flight
	power, and kinematics.
	\item
	\textbf{Acoustic telemetry}, used to monitor animal movement (mostly
	fish), or in some cases, effects of anthropogenic noise on animal
	behavior.
	\item
	\textbf{Experimental designs}, which involve analyzing behavioral and
	movement responses based on multiple stimuli, mostly on cattle and
	domestic animals.
	\item
	\textbf{Activity budgets}, investigating---mostly using telemetry
	data---the effect of environmental conditions on the time allocated to
	different activities.
	\item
	\textbf{Avian migration}, encompassing migration routes, orientation
	and flight strategies.
	\item
	\textbf{Sports}, consisting of motion analysis of sports players for
	better performance.
	\item
	\textbf{Human activity patterns}, mostly related to health and
	physical activity in children and adults, often sampled with
	accelerometers.
	\item
	\textbf{Breeding ecology}, involving space use and movement corridors
	during breeding seasons; mostly, but not exclusively on turtles and
	whales.
\end{enumerate}

While Sports and Human activity patterns are not strictly ecological
topics, they are an integral part of the literature studying movement
phenomena, and benefit from advances in movement ecology. For instance,
works on Sports have tried to understand why and how certain individual
and collective behaviors emerge in games, using principles from
ecological psychology (e.g.~\cite{Travassos2013}), focused on the interdependencies of
humans and their environments \cite{Barker1968}. Moreover, some studies related to
Human activity patterns were also inspired from animal studies
(e.g.~\cite{Wasenius2017}).

\begin{figure*}
	\includegraphics[width=17.8cm]{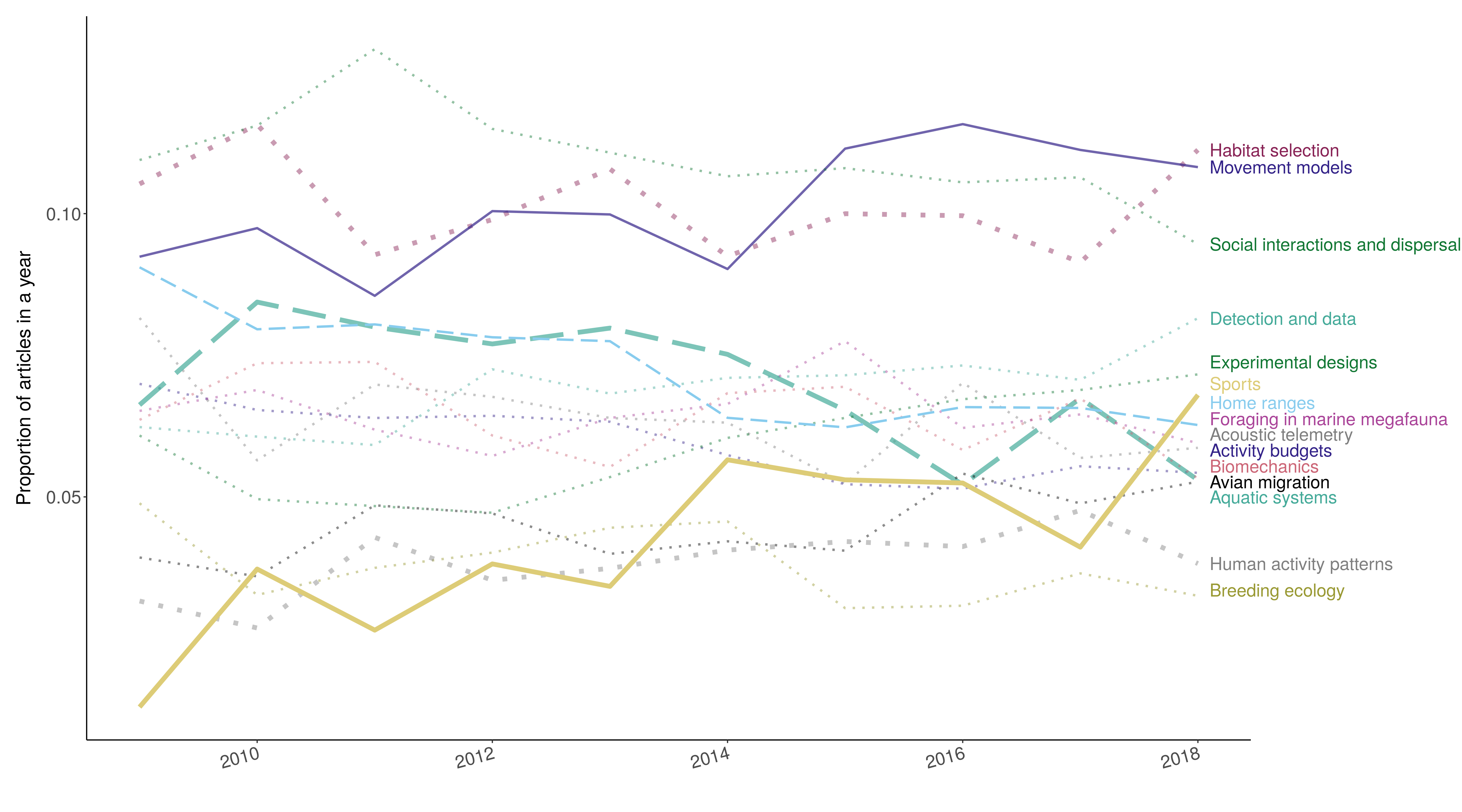}
	\caption{Time series of the relative prevalence of each topic every
		year. To improve readability, the topics with the most pronounced increases and decreases have been highlighted by continuous and dashed lines, respectively.}
	\label{fig:topics-ts}
\end{figure*}

Social interactions and dispersal, Movement models, and Habitat
selection remained the most relevant topics throughout the decade (Fig.
\ref{fig:topics-ts}). The prevalence of Home ranges studies decreased
over the years. In contrast, Sports has become a more recurrent topic in
the literature. The prevalence of the other topics has remained
relatively stable in time. The division into research topics has
revealed some distinction between marine and terrestrial realms, as four
topics pertained specifically to breeding or foraging ecology in marine
species.

\section*{The movement ecology framework}

\begin{figure}
	\includegraphics[width=250px]{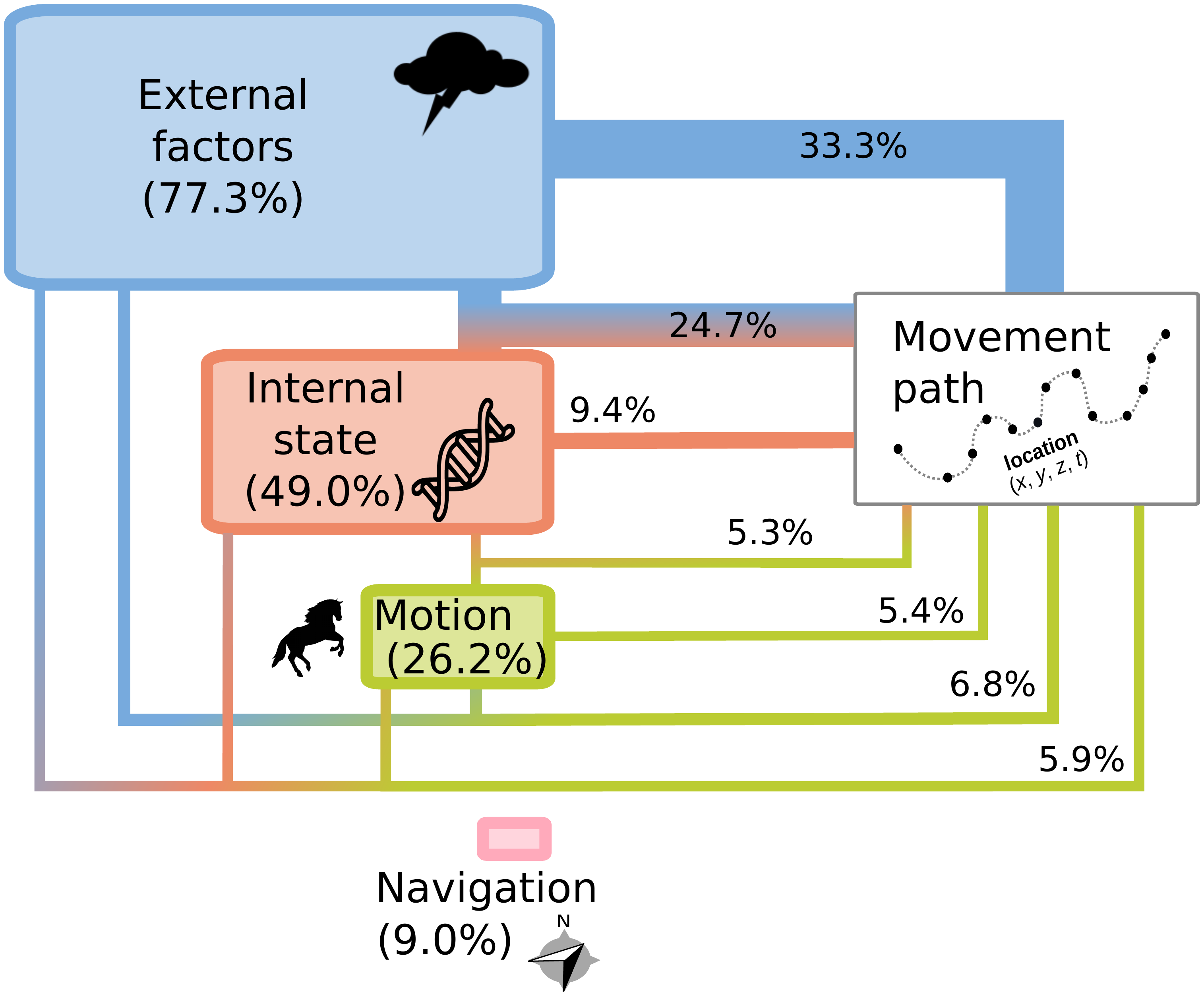}
	\caption{Representation of the components of the movement ecology framework and how much they were studied in the last decade: external factors, internal state, motion and navigation capacities, whose interactions result in the observed movement path. 
		The size of each component box is proportional to the percentage of papers (in parentheses) tackling them irrespectful of whether they are only about this component or in combination with another one. The latter is specified through the segments that join the components to the observed movement path. One fill color corresponds to papers that only studied one component, while two or more colors correspond to papers that tackled two or three components, respectively (the ones from those colors). The width of the segment is proportional to the percentage of papers that studied that combination (or single component). Only combinations corresponding to $>5\%$ of papers are shown; e.g. combinations involving navigation and papers studying navigation on its own had $<5\%$ of papers each therefore they are not shown in the graph.}
	\label{fig:framework}
\end{figure}

The MEF introduced by \cite{Nathan2008} consisted of four components: external factors
(i.e.~the set of environmental conditions that affect movement),
internal state (i.e.~the intrinsic factors affecting motivation and
readiness to move), navigation capacity (i.e.~the set of traits enabling
the individual to orient), and motion capacity (i.e.~the set of traits
enabling the individual to execute movement). The outcome of the
interactions between these four components would be the observed
movement path (plus observation errors). We found that, in the last
decade, most studies tackled movement in relation to external factors
(77\%), while a minority of them studied the three other components
(49\%, 26\%, and 9\%, for internal factors, motion, and navigation
capacity, respectively). For the most part, studies did not look into
interactions between these components, except for external factors with
internal states (25\% of the studies; Fig. \ref{fig:framework}). Quite
strikingly, this is the same overall pattern as in the decade before
(1999-2008; Appendix section 3.3.1).

\section*{Tools for movement ecology}

Technology has been a major driver of trends in data collection and
scientific publications in movement ecology. Past reviews have
highlighted an increase in the amount and variety of tracking devices,
which are becoming more affordable, with more efficient battery
capabilities, and reductions in size (see \cite{Thums2018}, \cite{Kays2015}, \cite{Harcourt2019}, and survey to
movement ecologists in Appendix section 4). Here, we categorized
tracking device observations as accelerometer, acoustic telemetry, body
condition measurements, encounter observations, GPS, light loggers,
pressure data, radar, radio telemetry, satellite, and video/image
(details of these categories and the analysis are in Appendix section
3.4).

\begin{figure}
	\includegraphics[width=420px]{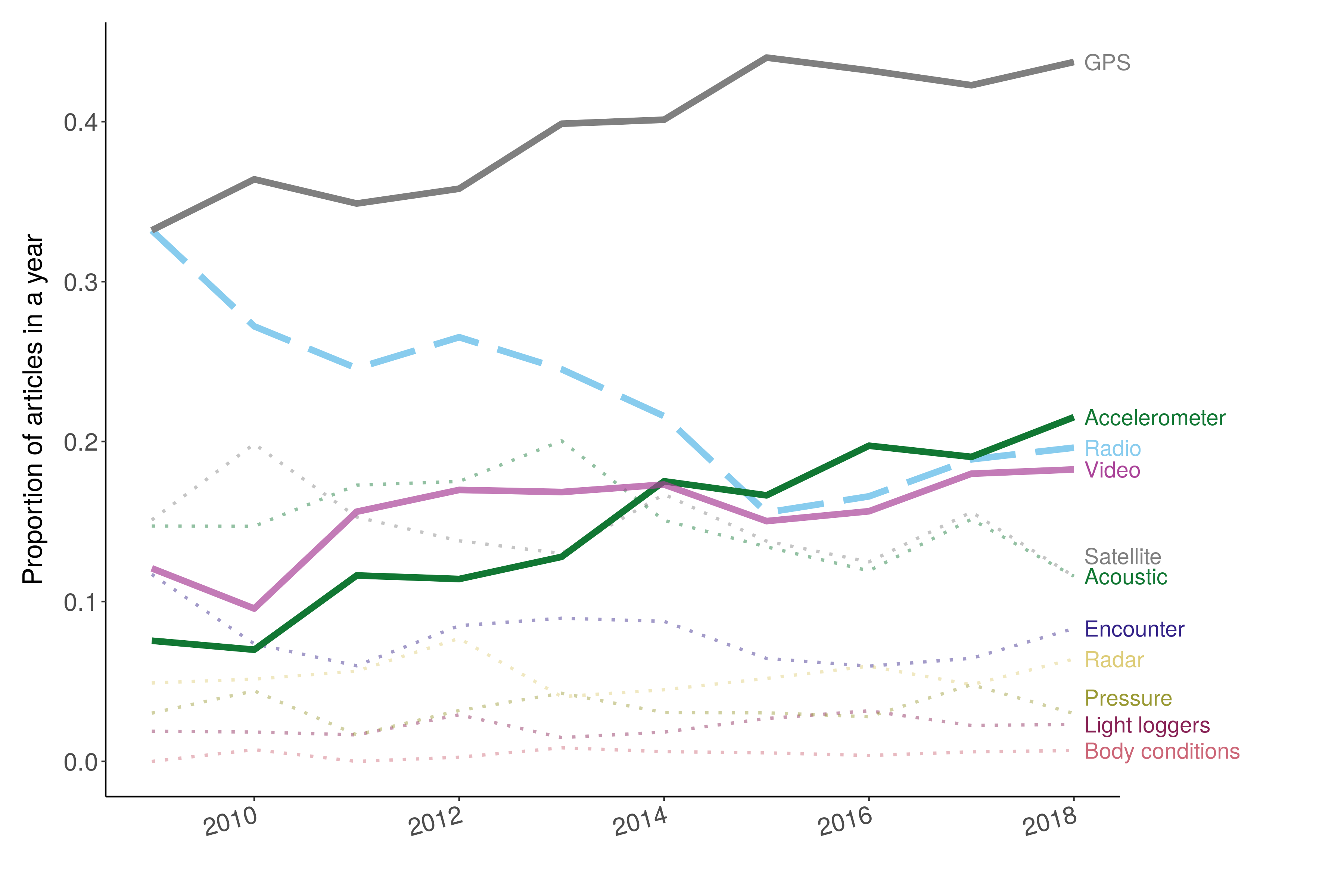}
	\caption{Proportion of papers in each year using each type of
		device. To improve readability, the devices with the most pronounced increases or decreases were highlighted by continuous and dashed lines, respectively.}
	\label{fig:devices-ts}
\end{figure}

Throughout the last decade, GPS has not only remained the most popular
device in movement studies, but its popularity in relation to other
methods has increased (Fig. \ref{fig:devices-ts}). This is likely due to
the development of cheaper, smaller and more efficient devices, which
make them a feasible option for small and medium-sized animals \cite{Kays2015}.
While in 2009 radio telemetry was as popular as GPS, later in the decade
GPS seems to have increasingly replaced radio telemetry \cite{Allan2018}, which has
been experiencing a substantial decrease in parallel. The use of
accelerometers and video is becoming more popular; the former allows for
finer spatio-temporal resolution movement data (Fig.
\ref{fig:devices-ts}), opening avenues to exploring physiological
aspects of movement like energy expenditure \cite{Wilson2019}, while the latter gives
us an animal's-eye view of its local environment, providing information
on visual cues used, foraging behavior and movement strategies \cite{Thiebault2014,Kane2015}.

\begin{figure}
	\includegraphics[width=420px]{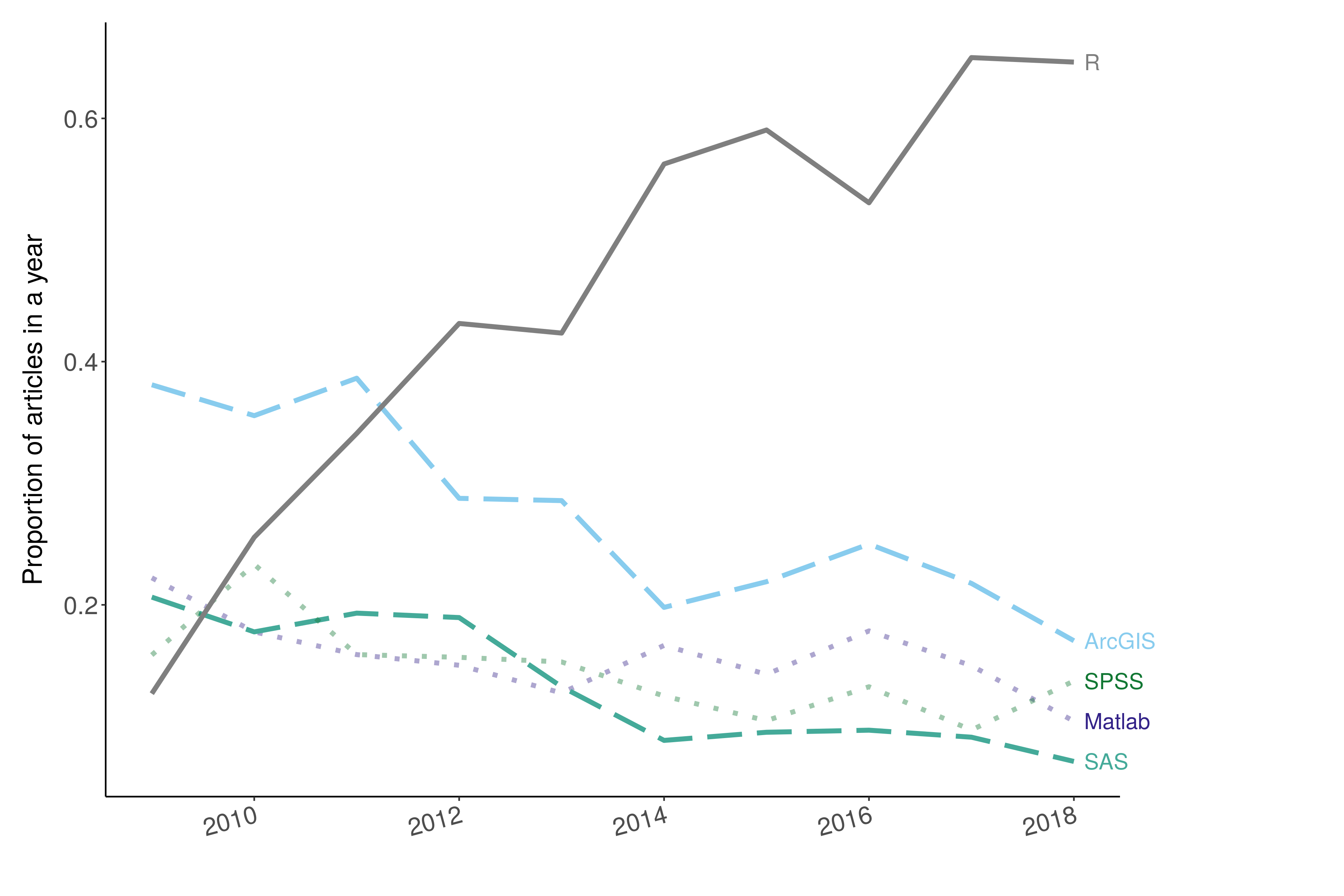}
	\caption{Proportion of papers of each year using each software package. Shown are the five most mentioned software packages. The ones with the most pronounced increases or decreases were highlighted by continuous and dashed lines, respectively.}
	\label{fig:software-ts}
\end{figure}

The increasing volume and diversity of movement data obtained through
these tracking devices require appropriate software tools for data
management, processing, and analysis \cite{Urbano2010,Joo2020}. We evaluated the use of
33 software packages (see section 3.5 of the Appendix for a full list),
and the five most popular through the decade were R, ArcGIS, Matlab,
SPSS and SAS, in that order. Among those, R experienced a constant and
strong growth in the last 10 years, while usage of all others
substantially decreased, making R an undisputed preference in the field
(Fig. \ref{fig:software-ts}).

In another study of ecology in general, the same pattern in reported R
usage was observed \cite{Lai2019}. According to both \cite{Lai2019} and this study, the
popularity of R ten years ago was low (used by \(> 10\%\) of the
papers), while the majority of articles published nowadays have reported
its use, indicating a homogenization of not only movement ecology but
ecology in general towards R. This success is likely due to the fact
that R offers a free software environment to program and create new
methods, share them, and improve them, facilitating transparency,
collaboration, and reproducibility \cite{Lowndes2017}, and at the same time it can be
extended with more than 50 specialized packages to process and analyze
movement \cite{Joo2020}. R also leverages other programming languages (e.g.~C,
python, Fortran, etc.) by allowing internal access to their use within
an R workflow (and R syntax).

In parallel with the development and improvement of tracking devices and
software, there has been substantial progress in the number and
sophistication of quantitative methods for the study of movement
(e.g.~\cite{Avgar2016,Seidel2018,Auger-methe2020}). We investigated the use of statistical methods
in the movement ecology literature (see Material and Methods, and
Appendix section 3.6, for more details). Most studies (68\%) used, at
the least, generic statistical methods (i.e.~with no explicit spatial,
temporal or social interaction component in its definition) such as
regression models. A smaller number (57\%) used at least one or more
specialized methods, i.e. movement, non-movement spatiotemporal
(e.g.~spatiotemporal geostatistics), spatial (e.g.~point process), time
series, and social analysis methods (e.g. social networks). Our analysis
reveals that researchers are not necessarily using movement-specific
techniques to analyze movement (only 33\% of the studies), and, in some
cases (42\%), not using spatial, temporal, or social analyses either.

\begin{figure}
	\includegraphics[width=420px]{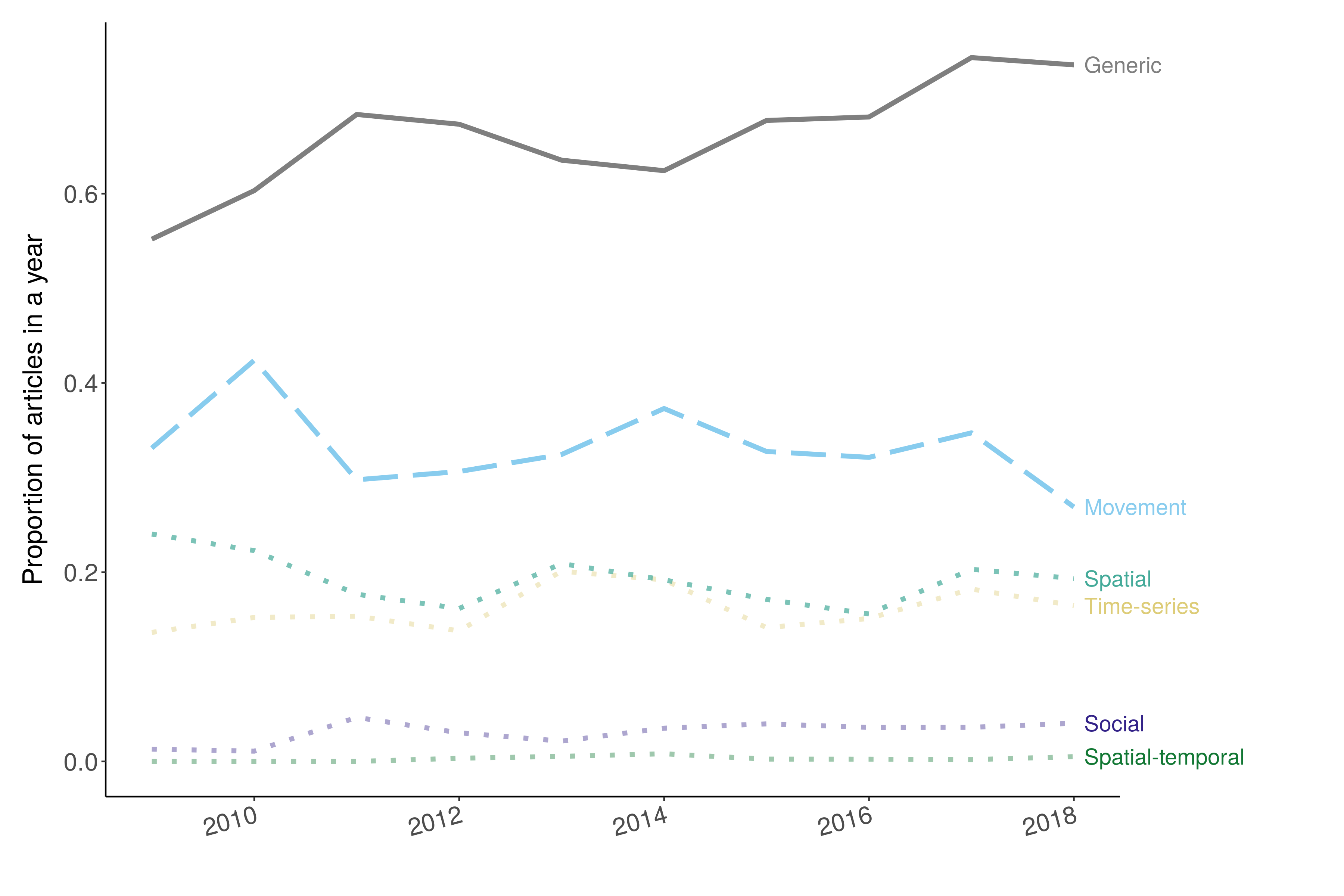}
	\caption{Proportion of papers of each year using each type of
		method. The types of method with the most pronounced increases or decreases were highlighted by continuous and dashed lines, respectively.}
	\label{fig:methods-ts}
\end{figure}

While the availability of movement data and associated software tools
and methods are increasing (see a summary list in \cite{Borger2016}), the proportion
of papers using movement-specific analytical methods does not show the
same pattern (Fig. \ref{fig:methods-ts}). Actually, the proportion of
usage of generic methods is increasing. In addition, and based on a
trigram analysis, we found that the most common methods were linear
mixed models (Table in section 3.6.1 of the Appendix).

This raises the question: why were the majority of papers not using
movement-specific methods? Certainly, not all studies require
movement-specific methods; the choice of method should depend on the
research question, assumptions, and data. Another reason for the use of
non-movement methods could be related to many scientists coming-of-age
in a time when autocorrelation in movement was considered a nuisance, or
they do not possess the quantitative skills necessary to use these
methods. Movement is a complex process, and in most cases, statistically
noisy, nonlinear, and spatially and temporally correlated (28).
Interdisciplinary work between ecologists and statisticians to
``decomplexify'' movement models (either making them more simple or
usable for different datasets and situations) may still be a challenge
to overcome \cite{Williams2020}. Moreover, as we intensify data collection and
processing, the use of movement models -- for a statistical
representation of movement and for population-level inference -- can be
expected to increase.

\section*{Open questions for the future of movement ecology}

Technology has undeniably been driving research in movement ecology in
the past decade. With access to numerous and diverse tracking data \cite{Baratchi2013},
and tools for data processing and analysis \cite{Borger2016,Joo2020}, researchers have
been able to sample spatiotemporal behavior and changes in physiology,
to investigate subjects like social interactions, habitat selection,
foraging behavior, physiological performance, and migration; topics that
were revealed by our text mining analysis. However, the technological
advances have not structurally changed the field of movement ecology:
none of those research topics are new, and, we have not moved towards
the integrative study of movement advised by the MEF.

The movement ecology framework was a revolutionary idea: an integrating
vision of the study of movement, represented by the interaction between
the four components of the framework. As argued by \cite{Nathan2008}, it is only
through combining different components of the framework that we can gain
a mechanistic understanding of movement, from the neurological and
physiological drivers to the life-history and evolutionary consequences.
While it has been recognized as a seminal, if not the most influential,
publication in the field (with \textgreater1000 citations according to
Web of Science), research in movement ecology has not translated into a
relative increase in publications addressing the internal state,
navigation, motion, or the interactions of the different components,
than in the decade before. The findings in \cite{Holyoak2008}, that the majority of
movement studies were ``simply measuring movement, documenting its
occurrence, or describing how it was influenced by the environment'',
still hold true. It may be for the simple reason that most researchers
in movement ecology are most likely ecologists. Ecology is inherently
focused on the environment; it is ``all the comprehensive science of the
relationship of the organism to the environment'' \cite{Haeckel1866}. It is logical
that, in a field mostly populated with ecologists, research questions
have been strongly related to the effect of external factors on animal
movement and behavior. However, movement ecology as a science should
pursue understanding of movement. The role of external factors on
movement can only be understood in conjunction with the internal state,
motion, and navigation capacities, even if these are closer to ethology,
biomechanics, and neuroscience, respectively, and studying these may not
seem as straightforward. Indeed, recent papers have addressed these
issues (e.g.~\cite{Goossens2020,McMahon2014,Morelle2014}) and opened perspectives for future
studies.

What does this mean for the future of movement ecology? As in every
field of science, there is a trade-off between data-driven and
ideas-driven research (see an analogous discussion for physical sciences
in \cite{Dyson2012}). It is likely that the developement of tagging devices,
statistical, and mathematical methods to describe patterns or model
processes will continue to shape our field (see survey results in
Appendix section 4). But will technological advances keep driving the
field more than movement concepts? Or will science be driven more by new
ideas to understand movement processes, inspiring the development and
use of particular technologies and analytical methods? Movement
ecologists need to decide where is the trade-off for their own research.
If we continue with the trend of the last two decades we will eventually
have to acknowledge the failure of the movement ecology framework as a
unifying paradigm of movement ecology.

Movement ecologists can rather choose to break this trend, and transform
the field into a more integrated science of movement. An integrated
approach to movement, as suggested by the MEF, would require truly
interdisciplinary efforts involving ecologists, biologists,
neuroscientists, physicists, and statisticians, among others \cite{Holyoak2008,Williams2020}.
It should aim at bridging the divide between human mobility research and
animal movement ecology, and between aquatic, terrestrial, and aerial
realms. These seemingly different fields may have questions and methods
in common \cite{Lowerre2019,Thums2018,Miller2019}; we could learn from each other and collaborate.

The path towards interdisciplinarity comes with many challenges \cite{Frodeman2017,Weingart2000}, concerning the researchers as individuals (particularly regarding
communication; \cite{Bracken2006,Marzano2006}) and as part of organizations that may not
have structures that encourage interdisciplinarity \cite{Brewer1999}, and, very
importantly, the difficulties of obtaining funds for interdisciplinary
research \cite{Bromham2016}. To face these challenges, movement ecologists should
direct systematic efforts towards interdisciplinarity. The field of
movement ecology would greatly benefit from exploring questions with
multiple and integrated approaches, novel and emerging movement
concepts. The path we choose to walk will be reflected on our research.

\section*{Materials and methods}

We selected scientific peer-reviewed papers in English that studied the
voluntary movement of one or more living individuals. We used Web of
Science (WoS) as a search engine. Since very few papers mention
``movement ecology'' in their abstracts, titles and keywords, we did not
use ``movement ecology'' as a search phrase. After a detailed testing
phase, we came up with search terms within four groups of words:
behavior, movement (e.g.~motion, moving), biologging (e.g.~telemetry,
gps) and individuals (e.g.~animal, human; we focused our efforts on
Animalia). To be qualified, papers needed to use words from at least 3
of these groups in their abstract, keywords or title. Also, papers
studying movement of objects other than whole organisms (e.g.~cell,
neuron), were filtered out. See more details on the search terms and
their quality control in section 2 of the Appendix. More than 8 thousand
(8,007) papers met our criteria. Results from the WoS (title, keywords,
abstracts, authors, DOI, etc.) on these 8,007 papers were downloaded and
used for most of the analyses. In addition, we used the
\texttt{fulltext} package \cite{Rfulltext} in \cite{R2018}, using Elsevier, Springer,
Scopus, Wiley, BMC, and PLOS one API keys to download full texts of
4,037 papers. Finally, using an automatic detection algorithm (see
section 2.3 of the Appendix for a description), 3,674 ``Materials and
Methods'' sections were extracted from this set of papers.

\subsection*{Topic analysis}

Topics were not defined \emph{a priori}. Instead, we fitted Latent
Dirichlet Allocation (LDA) models to the abstracts \cite{Blei2003}. LDAs are
basically three-level hierarchical Bayesian models for documents (in our
case, abstracts). Here we assumed that there was a fixed number of
latent or hidden topics behind the abstracts, and that the choice of
words in the abstracts were related to the topics the authors were
addressing. Thus, an abstract would have been composed of one or more
topics, and a topic would have been composed of a mixture of words. The
probability of a word appearing in an abstract depended on the topic the
abstract was adressing. Here we used the LDA model with variational EM
estimation \cite{Wainwright2008} implemented in the \texttt{topicmodels} package \cite{Grun2011}.
More details about the practical aspects of LDA modeling and a short
discussion on the number of topics can be found in section 3.1 of the
Appendix.

From the fitted LDA model, we obtained 1) for each topic, the
probability of the topic being referred to in each document (denoted by
\(\gamma\)), and 2) for each word, the probability of appearing in a
document given the presence of a certain topic (denoted by \(\beta\)).
The \(\beta\) values were thus a proxy of the importance of a word in a
topic. They were used to interpret and label each topic, and to create
wordclouds for each topic, shown in section 3.1 of the Appendix. The sum
of \(\gamma\) values for each topic served as proxies of the
``prevalence'' of the topic relative to all other topics and were used
to rank them.

\subsection*{Taxonomical identification}

To identify the taxonomy of the individuals studied in the papers, the
ITIS (Integrated Taxonomic Information System) database
(\url{https://www.itis.gov/citation.html}) was used to detect names of
any animal species (kingdom Animalia) that were mentioned in the
abstracts, titles and keywords. We screened these sections for latin and
common (i.e., vernacular) names of species (both singular and plural),
as well as common names of higher taxonomic levels such as orders and
families. After having identified any taxon mentioned in a paper, we
summarized taxa at the Class level (except for superclasses Osteichthyes
and Chondrichthyes which we merged into a single group labeled Fish, and
for classes within the phylum Mollusca and the subphylum Crustacea which
we considered collectively). Thus, each paper was classified as focusing
on one or more class-like groups; for example - mammals, birds, insects,
etc. For the purpose of our analysis, we kept humans as a separate
category and did not count them within Class Mammalia. See section 3.2
of the Appendix for more details.

\subsection*{Framework and tools}

To assess the study of the different components of the movement ecology
framework, we built what we call here a ``dictionary''. A dictionary is
composed of words and their meanings. Here, the words of interest were
the components of the framework (i.e.~internal state, external factor,
motion, and navigation), and their meanings were the terms potentially
used in the abstracts to refer to the study of each component. For
example, terms like ``memory'', ``sensory information'', ``path
integration'', or ``orientation'' were used to identify the study of
navigation. Similarly, the devices, software, and statistical methods
used were also assessed through dictionary approaches. More details,
including quality control of the dictionaries, in sections 3.3 to 3.6 of
the Appendix.

\subsection*{Access to data and codes}

We provide details on all data processing and analyses at
\url{https://rociojoo.github.io/mov-eco-review/}, from descriptions of
word search on Web of Science and scripts to download the papers, up to
the codes to reproduce every single plot in this manuscript. The
website, hosted in the \url{https://github.com/rociojoo/mov-eco-review}
repository, works as an online Appendix to this manuscript. The authors can be directly
contacted for further development and questions about the dataset, which
has not been released to respect Text and Data Mining rights of the
publishers.

\section*{Acknowledgments}

The authors would like to thank Susana Clusella-Trullas for fruitful
exchanges about internal states and physiology. Trey Shelton, from UF
library, gave advice and guidance about TDM rights and obtaining APIs
from publishers, which was very useful at early stages of this study. We
are also grateful to Luis Cajachahua Espinoza, for his help to explore
scrapping possibilities at the very beginning of this work. RJ, MEB,
TAC, SCP and MB were funded by a Human Frontier Science Program Young
Investigator Grant (SeabirdSound; RGY0072/2017).

\newpage

 \bibliographystyle{abbrv}
 

\end{document}